\documentclass[12pt]{article}
\usepackage{epsfig}
\usepackage{amssymb}
\usepackage{amsmath}
\usepackage{amsfonts}
\usepackage{graphicx}
\usepackage{mathrsfs}
\usepackage[dvips]{color}
\usepackage{multirow}

\newcommand{\R}{\mathbb{R}}
\newcommand{\C}{\mathbb{C}}

\newcommand{\fp}{\mathfrak{p}}

\newcommand{\fz}{\mathfrak{z}}

\newcommand{\cH}{\mathcal{H}}

\newcommand{\be}{\begin{equation}}
\newcommand{\ee}{\end{equation}}
\newcommand{\bea}{\begin{eqnarray}}
\newcommand{\eea}{\end{eqnarray}}
\newcommand{\nn}{\nonumber}

\newcommand{\ed}{\end{document}}

\newcommand{\bi}{\begin{itemize}}
\newcommand{\ei}{\end{itemize}}

\newcommand{\bce}{\begin{center}}
\newcommand{\ece}{\end{center}}

\newcommand{\RE}{{\rm Re}}
\newcommand{\IM}{{\rm Im}}

\begin{document}

\title{A Differential Integrability Condition for Two-Dimensional Hamiltonian Systems}

\author{Ali~Mostafazadeh\thanks{E-mail address:
amostafazadeh@ku.edu.tr, Phone: +90 212 338 1462, Fax: +90 212 338
1559}
\\
Department of Mathematics, Ko\c{c} University,\\
34450 Sar{\i}yer, Istanbul, Turkey}
\date{ }
\maketitle

\begin{abstract}
We review, restate, and prove a result due to Kaushal and Korsch [Phys.\ Lett.~A {\bf 276}, 47 (2000)] on the complete integrability of two-dimensional Hamiltonian systems whose Hamiltonian satisfies a set of four linear second order partial differential equations. In particular, we show that a two-dimensional Hamiltonian system is completely integrable, if the Hamiltonian has the form $H=T+V$ where $V$ and $T$ are respectively harmonic functions of the generalized coordinates and the associated momenta.
\medskip

\end{abstract}

\maketitle

The study of integrable Hamiltonian systems has been a subject of active research since the nineteenth century. Despite the existence of an extensive literature on the subject, there is no well-known practical test of complete integrability of a given Hamiltonian system even in two dimensions. The present work elaborates on a simple-to-check sufficient condition for the complete integrability of such Hamiltonian systems. This condition, which is originally outlined in Ref.~\cite{KK-2000}, identifies a special class of completely integrable Hamiltonian systems in two dimensions. The following is a precise statement of this result.
	\begin{itemize}
	\item[~]{\bf Theorem~1} Let $S$ be a Hamiltonian system with phase space $\R^4$ and a twice-continuously-differentiable time-independent Hamiltonian $H:\R^4\to\R$ that satisfies
	\begin{align}
	&\frac{\partial^2 H}{\partial x_1^2}+\frac{\partial^2 H}{\partial x_2^2}=0, &&
	\frac{\partial^2 H}{\partial p_1^2}+\frac{\partial^2 H}{\partial p_2^2}=0,
	\label{condi1a}\\
	&\frac{\partial^2 H}{\partial x_1\partial p_2}+
	\frac{\partial^2 H}{\partial x_2\partial p_1}=0, &&
	\frac{\partial^2 H}{\partial x_1\partial p_1}-
	\frac{\partial^2 H}{\partial x_2\partial p_2}=0,
	\label{condi1b}
	\end{align}
where $H=H(x_1,p_1,x_2,p_2)$, and $(x_1,p_1)$ and $(x_2,p_2)$ are conjugate coordinate-momentum pairs. Then $S$ is completely integrable.
	\end{itemize}
Before giving a proof of this theorem, we wish to motivate its statement and the idea of its proof.

Let $\fz:=x+i y$ and $\fp:=p+i q$ be a pair of complex variables, where $x,y,p,q\in\R$, and $\cH:\C^2\to\C$ be an analytic complex-valued function of $(\fz,\fp)$ with real and imaginary parts $u$ and $v$, so that
	\be
	u(x,y,p,q):=\RE[\cH(x+i y,p+i q)],~~~~~~
	v(x,y,p,q):=\IM[\cH(x+i y,p+i q)].
	\label{u-v}
	\ee
We can express the condition of the analyticity of $\cH$ as the Cauchy-Riemann relations:
	\begin{align}
	&\frac{\partial u}{\partial x}=\frac{\partial v}{\partial y},
	&& \frac{\partial u}{\partial y}=-\frac{\partial v}{\partial x},
	&&\frac{\partial u}{\partial p}=\frac{\partial v}{\partial q},
	&& \frac{\partial u}{\partial q}=-\frac{\partial v}{\partial p}.
	\label{CR}
	\end{align}
These in turn imply
	\begin{align}
	&\frac{\partial^2 u}{\partial x^2}+\frac{\partial^2 u}{\partial y^2}=0, &&
	\frac{\partial^2 u}{\partial p^2}+\frac{\partial^2 u}{\partial q^2}=0,
	&&\frac{\partial^2 u}{\partial x\partial p}+\frac{\partial^2 u}{\partial y\partial q}=0, &&
	\frac{\partial^2 u}{\partial x\partial q}-\frac{\partial^2 u}{\partial y\partial p}=0.
	\label{condi-3}
	\end{align}
Refs.~\cite{KK-2000,pla-2006} show that the complexified Hamilton equations,
	\begin{align}
	&\frac{d\fz}{dt}=\frac{\partial\cH}{\partial\fp}, &&
	\frac{d\fp}{dt}=-\frac{\partial\cH}{\partial\fz},
	\label{complex-HE}
	\end{align}
define a completely integrable Hamiltonian system in the phase space $\R^4$ whose dynamics
may be determined using the standard Hamilton equations with $u$ serving as the Hamiltonian and $v$ acting as a constant of motion. It is important to note that in order to establish this result one needs to use an appropriate symplectic structure (which differs from but elbeit isomorphic to the standard symplectic structure) on $\R^4$ and the associated Darboux coordinates for the problem \cite{pla-2006,CM-2007}. The latter can be taken to be the conjugate coordinate-momentum pairs $(x,p)$ and $(q,y)$, i.e.,  $q$ and $y$ must be viewed as coordinate and momentum variables, respectively \cite{Xavier}.

The statement and proof of Theorem~1 relies on the fact that Eqs.~(\ref{condi1a}) and (\ref{condi1b}), that were initially derived in \cite{KK-2000}, can be used to identify $H$ with the real part of an analytic function $\cH:\C^2\to\C$ such that the Hamiltonian system determined by $H$ coincides with the one defined by (\ref{complex-HE}). In other words, not only they are necessary but also sufficient for the complete integrability of the system. The following is a direct and explicit proof of this assertion. 
\begin{itemize}
\item[]{\bf Proof of Theorem~1:} Let $x:=x_1$, $y:=x_2$, $p:=p_1$, $q:=-p_2$, and
	\begin{align}
	&u(x,y,p,q):=H(x,p,y,-q).
	\label{u=H}
	\end{align}
Then (\ref{condi1a}) and (\ref{condi1b})  are equivalent to  (\ref{condi-3}), and if we view  (\ref{CR}) as a system of differential equations for  $v$, then (\ref{condi-3}) becomes the integrability condition for these equations. This shows that there is a solution $v$ of (\ref{CR})
for the case that $u$ is given by (\ref{u=H}). Let $I:\R^4\to \R$ be defined by
	\be
	I(x_1,p_1,x_2,p_2):=v(x_1,x_2,p_1,-p_2),
	\label{tI=}
	\ee
so that
	\be
	v(x,y,p,q)=I(x,p,y,-q).
	\label{v=tI}
	\ee
We can use (\ref{u=H}) and (\ref{v=tI}) to express (\ref{CR}) in the form
	\begin{align}
	&\frac{\partial H}{\partial x_1}=\frac{\partial I}{\partial x_2},
	&&\frac{\partial H}{\partial x_2}=-\frac{\partial I}{\partial x_1},
	&&\frac{\partial H}{\partial p_1}=-\frac{\partial I}{\partial p_2},
	&&\frac{\partial H}{\partial p_2}=\frac{\partial I}{\partial p_1}.
	\label{condi-4}
	\end{align}
These in turn imply $\{ H, I\}=0$, where $\{\cdot,\cdot\}$ is the Poisson bracket,
	\be
	\{ H, I\}:=\sum_{k=1}^2\left(
	\frac{\partial  H}{\partial x_k}\frac{\partial  I}{\partial p_k}
	-\frac{\partial I}{\partial x_k}\frac{\partial H}{\partial p_2}\right).
	\nn
	\ee
This identifies $I$ with a constant of motion for the system $S$. Next, we prove that $I$ is functionally independent of $H$. Assume by contradiction that $I=F(H)$ for some real-valued differentiable function $F:\R\to\R$. Then for all $z\in\{x_1,p_1,x_2,p_2\}$,
	\begin{align}
	\frac{\partial I}{\partial  z}=
	F'\:\frac{\partial H}{\partial z},
	\label{indep}
	\end{align}
where $F'$ stands for the derivative of $F$. Substituting (\ref{indep}) in (\ref{condi-4}), we find
	\begin{align*}
	&\frac{\partial H}{\partial x_1}=F'\:\frac{\partial H}{\partial x_2},
	&&\frac{\partial H}{\partial x_2}=-F'\:\frac{\partial H}{\partial x_1},	
	&&\frac{\partial H}{\partial p_1}=-F'\:\frac{\partial H}{\partial\tilde p_2},
	&&\frac{\partial H}{\partial p_2}=F'\:\frac{\partial H}{\partial\tilde p_1}.
	\end{align*}
We can use these equations to establish
	\[\left({F'}^2+1\right)\frac{\partial H}{\partial x_1}=
	\left({F'}^2+1\right)\frac{\partial H}{\partial x_2}=
	\left({F'}^2+1\right)\frac{\partial H}{\partial p_1}=
	\left({F'}^2+1\right)\frac{\partial H}{\partial p_2}=0.\]
Clearly these cannot be satisfied unless $H$ is a constant function, which is definitely not the case. This completes the proof that $I$ is a constant of motion that is functionally independent of $H$. Therefore, according to Liouville's Integrability Theorem, $S$ is completely integrable.~~$\square$
\end{itemize}

The following is a straightforward consequence of Theorem~1.
	\begin{itemize}
\item[]{\bf Theorem~2}  (Harmonic Integrability) Let $T:\R^2\to\R$, $V:\R^2\to\R$, and $H:\R^4\to\R$ be  twice-continuously-differentiable functions such that $H$ serves as the Hamiltonian for a classical system $S$ with phase space $\R^4$ and has the form
	\be
	H(x_1,p_1,x_2,p_2)=T(p_1,p_2)+V(x_1,x_2).
	\label{sep}
	\ee
Then $S$ is completely integrable, if $T$ and $V$ are harmonic functions.
\item[]{\bf Proof:} For the Hamiltonians of the form~(\ref{sep}), (\ref{condi1b}) is trivially satisfied and (\ref{condi1a}) is equivalent to the condition that $T$ and $V$ are harmonic functions. Therefore, according to Theorem~1, $S$ is completely integrable, if $T$ and $V$ are harmonic functions.~~$\square$
	\end{itemize}

Given a Hamiltonian $H$ that fulfils (\ref{condi1a}) and (\ref{condi1b}), we can use (\ref{condi-4}) to construct an independent constant of motion $I$. In order to demonstrate this construction, we examine three concrete examples. In the first two of these
$H$ satisfies the conditions of Theorem~2. In all of them, $c$ is an arbitrary real constant.\\[6pt]
\noindent {\bf Example~1}: Consider the Hamiltonian $H:=\frac{1}{2}\left(p_1^2-p_2^2-x_1^2+x_2^2\right)$ that is clearly of the form (\ref{sep}) for a pair of harmonic functions $T$ and $V$. We can write (\ref{condi-4}) as
	\begin{align}
	&\frac{\partial I}{\partial x_2}=-x_1,
	&&\frac{\partial I}{\partial x_1}=-x_2,
	&& \frac{\partial I}{\partial p_2}=-p_1,
	&&\frac{\partial I}{\partial p_1}=-p_2,
	\label{condi-e1}
	\end{align}
which we can easily integrate to find $I=-(x_1x_2+p_1p_2)+c$.\\[6pt]
\noindent {\bf Example~2}: The Hamiltonian $H:=e^{p_1}\cos p_2+e^{-x_1}\sin x_2$ also fulfils the hypothesis of Theorem~2. For this Hamiltonian, (\ref{condi-4}) takes the form
	\begin{align}
	&\frac{\partial I}{\partial x_2}=-e^{-x_1}\sin x_2,
	&&\frac{\partial I}{\partial x_1}=-e^{-x_1}\cos x_2,
	&& \frac{\partial I}{\partial p_2}=-e^{p_1}\cos p_2,
	&&\frac{\partial I}{\partial p_1}=-e^{p_1}\sin p_2.
	\label{condi-e1}
	\end{align}
Again we can integrate these equations and obtain $I=e^{-x_1}\cos x_2-e^{p_1}\sin p_2+c$.\\[6pt]
\noindent {\bf Example~3}: Consider the Hamiltonian
	\[H:=\frac{1}{4}\left(x_1^4-6x_1^2x_2^2+x_2^4\right)
	\left(p_1^4-6p_1^2p_2^2+p_2^4\right)+
	4x_1x_2\left(x_1^2-x_2^2\right)p_1p_2\left(p_1^2-p_2^2\right).\]
It is not difficult to check that it satisfies (\ref{condi1a}) and (\ref{condi1b}), but that it cannot be expressed in the form (\ref{sep}). Integrating (\ref{condi-4}) for this choice of $H$ gives
	\[ I=\left(x_1p_1+x_2p_2\right)\left(x_2p_1+x_1p_2\right)
	\left[(x_1+x_2)p_1-(x_1-x_2)p_2\right]
	\left[(x_1-x_2)p_1+(x_1+x_2)p_2\right]+c.\]

\subsection*{Acknowledgments} This work has been supported by the Turkish Academy of Sciences (T\"UBA).

\ed

\bibitem{other} R.~S.~Kaushal and S.~Singh, ``Construction of complex invariants for classical dynamical systems,'' Ann.\ Phys.\ (N.Y.) {\bf 288}, 253-276 (2001);

A.~V.~Smilga, ``Cryptogauge symmetry and cryptoghosts for crypto-Hermitian Hamiltonians,''
J.~Phys.~A {\bf 41}, 244026 (2008).